\documentclass[prb,twocolumn,aps,showpacs,superscriptaddress,floatfix] {revtex4}

\usepackage{graphicx}
\usepackage{latexsym}
\usepackage{amsmath}
\usepackage{color}
\usepackage[colorlinks,bookmarks=false,citecolor=blue,
linkcolor=red,urlcolor=blue]{hyperref}

\begin{document}

\title{
Degeneracy and ordering of the non-coplanar phase of the classical
bilinear-biquadratic Heisenberg model on the triangular lattice }

\author{S.~E. Korshunov}
\affiliation{L.~D. Landau Institute for Theoretical Physics RAS, 142432
Chernogolovka, Russia}
\author{F. Mila}
\affiliation{Institute of Theoretical Physics, Ecole Polytechnique
F\'ed\'erale de Lausanne, CH-1015
Lausanne, Switzerland}
\author{K. Penc}
\affiliation{Research Institute for Solid State Physics and
Optics, H-1525 Budapest, P.O. Box 49, Hungary}

\date{February 15, 2012}


\begin{abstract}
We investigate the zero-temperature behavior of the classical Heisenberg
model on the triangular lattice in which the competition between exchange
interactions of different orders favors a relative angle between
neighboring spins $\Phi\in(0,2\pi/3)$. In this situation, the ground
states are noncoplanar and have an infinite discrete degeneracy.
In the generic case, i.e. when $\Phi\neq\pi/2,\arccos(-1/3)$, the ground
state manifold is in one to one correspondence (up to a global rotation)
with the set of non-crossing loop coverings of the three equivalent
honeycomb sublattices into which the bonds of the triangular lattice can
be partitioned. This allows one to identify the order parameter space as
an infinite Cayley tree with coordination number 3. Building on the
duality between a similar loop model and the ferromagnetic O(3) model on
the honeycomb lattice, we argue that a typical ground state should have
long-range order in terms of spin orientation. This conclusion is further
supported by the comparison with the four-state antiferromagnetic Potts
model [describing the $\Phi=\arccos(-1/3)$ case], which at zero
temperature is critical and in terms of the solid-on-solid representation
is located exactly at the point of roughening transition. At $\Phi\neq
\arccos(-1/3)$ an additional constraint appears, whose presence drives the
system into an ordered phase (unless $\Phi=\pi/2$, when another constraint
is removed and the model becomes trivially exactly solvable).
\end{abstract}

\maketitle

\section{Introduction}

The recent discovery of the compound NiGaS$_4$, and the suggestion that it
might be a spin liquid or a spin nematic, \cite{nakatsuji} has revived the
interest in the Heisenberg model with bilinear and biquadratic
interactions defined by the Hamiltonian
\begin{equation}
  \mathcal{H} =  \sum_{(jj')} J_1 {\bf S}_{j} \cdot {\bf S}_{j'}
  + J_2 \left({\bf S}_{j} \cdot {\bf S}_{j'}\right)^2\;,    \label{eq:Hbb}
\end{equation}
where $({j}{j}')$ stands for nearest neighbor pairs on a lattice. In one
dimension, the spin-1 case has been very thoroughly investigated following
Haldane's prediction that the pure Heisenberg model is
gapped,\cite{haldane} and a rather complete picture of its properties has
emerged both in zero  (see e.g.
Refs.~[\onlinecite{PRL_AKLT,fath1991,Schollwock1996}] and references
therein) and in finite magnetic field.\cite{parkinson,okunishi1,fath,
manmana}

In two dimensions, most of the attention has also been devoted to the
spin-1 case, and some trends have emerged from the investigation of this
model on the triangular and square lattices.
\cite{kawashima,arikawa,LMP,toth} With the standard notation $J_1=J\cos
\theta$, $J_2=J \sin \theta$, the phase diagram as a function of $\theta$
consists of four main phases: an antiferromagnetic phase (with two
sublattices on the square lattice and three sublattices on the triangular
lattice) around $\theta=0$, followed counter-clockwise by a
three-sublattice antiferroquadrupolar phase up to $\theta=\pi/2$, a
ferromagnetic phase around $\theta=\pi$,  and finally a ferroquadrupolar
phase that persists until the antiferromagnetic order sets in.

In comparison, little attention has been paid to the classical limit,
where spins are classical vectors of length $1$. The main reason lies in
the fact that, unlike for spin-1/2 models, in the spin-1 case the
classical limit is not equivalent to the Hartree approximation (in which
the ground state is approximated by a product of local wave functions)
since local wave functions are not necessarily purely magnetic but can
also describe quadrupolar states.

In the classical limit, the most unusual feature introduced by the
presence of the biquadratic exchange is the possibility (realized in the
parameter range \makebox{$-2J_2<J_1<2J_2$} with $J_2>0$) to have the
minimum of the interaction of two spins in a {\em noncollinear}
configuration in which they make an angle
\begin{equation}
 \Phi=\arccos\left(-\frac{J_1}{2 J_2}\right)\in(0,\pi)
 \label{eq:Phi}
\end{equation}
with respect to each other, while the overall orientation of the spins is
arbitrary. \cite{LGT,KY} Such an interaction may be called a {\it
rotationally invariant spin-canting interaction}. It should not be
confused with the Dzyaloshinskii-Moriya interaction, which also induces a
canting between a pair of spins but forces them to lie in a specific
plane.

Note that the physics will be essentially the same for any rotationally
invariant canting interaction between pairs of spins that leads to the
same angle $\Phi$, for instance interactions that include higher powers of
${\bf S}_j \cdot {\bf S}_{j'}$, in which case the relation (\ref{eq:Phi})
would have to be replaced by a more complex one. For this reason we prefer
below to characterize the interaction in terms of the angle $\Phi$ and not
in terms of the parameter $\theta$. The classical XY model with
rotationally invariant spin-canting interaction on the square lattice has
been investigated by Lee {\em et al.} \cite{LGT} and the classical
Heisenberg model on the triangular lattice by Kawamura and Yamamoto.
\cite{KY}

\begin{figure}[b]
$~$
\includegraphics[width=70mm]{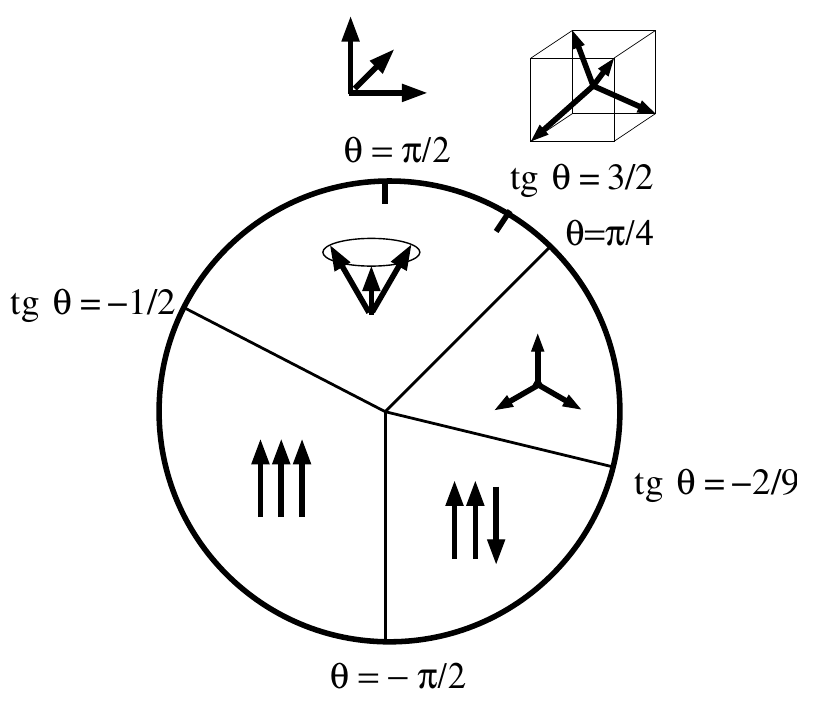}
\caption[fig-v] {Phase diagram of the classical bilinear-biquadratic
Heisenberg model on the triangular lattice.}
\label{fig_phase_diagram_circle}
\end{figure}

According to Ref.~\onlinecite{KY}, 
the zero-temperature phase diagram of the classical version of model
{\ref{eq:Hbb}} on the triangular lattice includes four phases (see
Fig.~\ref{fig_phase_diagram_circle}): a ferromagnetic phase for
$\pi-\arctan(1/2)<\theta<3\pi/2$, a nematic-like phase with collinear
orientation of spins for $-\pi/2<\theta<-\arctan(2/9)$, a three-sublattice
antiferromagnetic phase for $-\arctan(2/9)<\theta<\pi/4$, and a phase with
non-coplanar orientations of spins for
\makebox{$\pi/4<\theta<\pi-\arctan(1/2)$}. This phase diagram is
reminiscent of the spin-1 phase diagram, but the non-degenerate ferro- and
antiferroquadrupolar phases of the spin-1 case are replaced by highly
degenerate magnetic phases. The nematic-like phase is thoroughly discussed
in Ref.~\onlinecite{KY}, with the conclusion that it has the same
degeneracy and correlations as the antiferromagnetic Ising model on the
triangular lattice.

The discussion of the non-coplanar phase is much more sketchy however. The
authors point out that there is a degeneracy due to the two possible
chiralities of a local umbrella [formed on each triangular plaquette when
$\Phi\in (0,2\pi/3)$], but no attempt has been made to identify the
structure of the ground state manifold or to study the consequences of the
degeneracy on the spin correlations in a typical ground state. In the
present paper, we construct the complete classification of the ground
states which allows us to conclude that the non-coplanar phase is a
non-trivial one even at zero temperature because the total number of
ground states grows exponentially with the area of the system, as in
various versions of the antiferromagnetic (AF) Potts model and other
frustrated models with a finite residual entropy per site.
\cite{Suto,KB,St,B70,HR,KSS} In situations like this, it is not evident
{\it a priori} whether the degeneracy leads at zero temperature to the
disordering of the system or to an algebraic decay of correlations, or
whether long-range order is nevertheless present.

Interestingly enough, in the two-dimensional antiferromagnets known to
have an extensive residual entropy, all three scenarios are realized. In
particular, both the AF Ising model on the {\em kagom\'{e}} lattice
\cite{Suto} and the three-state AF Potts model on the square lattice
\cite{KB} at zero temperature are disordered, the AF Ising model on the
triangular lattice \cite{St} and the three-state AF Potts model on the
{\em kagom\'{e}} lattice \cite{B70,HR} are critical, whereas the
three-state AF Potts model on the dice lattice has a genuine long-range
order. \cite{KSS} The main aim of our paper is to establish which of the
three scenarios is realized in the present case.

The paper is organized as follows. Sec. \ref{GrStates} is devoted to a
complete classification of the ground states of the model in terms of the
nonintersecting closed loops living on the bonds of the triangular
lattice, which allows us to redefine the problem as a loop model. In Sec.
\ref{O(n)} we show that the order-parameter manifold of this loop model
has the same topology as that of the $O(n)$ model with $n=3$, which
suggests that they belong to the same universality class and, therefore,
that both of them are in the ordered phase. In Sec. \ref{Potts}, the same
conclusion is confirmed by establishing a relation with the four-state AF
Potts model on the same lattice and by relying on some known properties of
this Potts model. For completeness, Sec. \ref{Pi/2} briefly discusses the
case \makebox{$\Phi=\pi/2$}, where the model is trivially exactly
solvable, and Sec. \ref{Concl} summarizes the results. Appendix
\ref{RProblem} is devoted to the analysis of a simplified problem
corresponding to freezing the spins on one of the three sublattices in a
perfectly ordered state. It also includes estimates from below and from
above of the residual entropy of the full problem.

\section{Ground state classification and loop model\label{GrStates}}

We assume that classical spins ${\bf S}_j$ (three-dimensional vectors of
unit length) are located at the sites ${j}$ of the triangular lattice and
that the energy of the interaction of two neighboring spins $E({\bf
S},{\bf S}')$ depends only on the angle between ${\bf S}$ and ${\bf S}'$
and is minimal when this angle is equal to $\Phi \in (0,2\pi/3)$. In terms
of the coupling constants $J_{1}$ and $J_2$ introduced in
Eq.~(\ref{eq:Hbb}) this corresponds to \makebox{$-2J_2<J_1<J_2$} (see
Eq.~\ref{eq:Phi}). For such values of $\Phi$, it turns out to be possible
to minimize $E({\bf S}_j,{\bf S}_{j'})$ simultaneously for all pairs of
neighboring spins $({jj}')$, which means that the system is not
frustrated. However, as we shall see, it is very highly degenerate.

The simplest ground states of the model can be constructed by choosing
three unit vectors ${\bf S}^A$, ${\bf S}^B$ and ${\bf S}^C$ which form
angles $\Phi$ with respect to each other. After that, one can partition
the triangular lattice into three triangular sublattices (denoted below A,
B, and C) and set all spins on sublattice A to ${\bf S}^A$, on sublattice
B to ${\bf S}^B$ and on the third one to ${\bf S}^C$. In such a way, one
obtains a periodic ground state which has a three-sublattice structure.
\cite{KY} Below we call such ground states regular states.

The family of regular states is characterized by an $SO(3)\times Z_2$
degeneracy, where the group $SO(3)$ is related to the possibility of
simultaneously rotating all spins and the group $Z_2$ to the possibility
of choosing the sign of the mixed product of the three spins located in
the corners of a given plaquette to be either positive or negative.
However, it is clear that the family of ground states is much wider than
the set of regular states, and that, in addition to these two
symmetry-induced degeneracies, it possesses also an infinite set of
discrete degeneracies not related to global symmetries. Indeed, for any
pair of spins  ${\bf S}^{A}$ and ${\bf S}^{B}$ forming an angle $\Phi \in
(0,2\pi/3)$ with respect to each other there are always two possibilities
to choose the orientation of the third spin in such a way that it forms
the same angle $\Phi$ both with ${\bf S}^A$ and with ${\bf S}^B$. These
two orientations are mirror images of each other with respect to the plane
formed by  ${\bf S}^A$ and ${\bf S}^B$. Since in a regular state all
plaquettes contain the same triad of spins (${\bf S}^A$, ${\bf S}^B$ and
${\bf S}^C$), it is clear that in such a state one can always flip an
arbitrary spin into its mirror image without increasing the energy of the
system.

Flipping one spin leads to a new ground state in which six neighboring
triangular plaquettes (surrounding the flipped spin) form a domain of a
different regular state. In order to obtain a larger domain, one has to
flip other spins belonging to the
same sublattice. As soon as a domain of another regular state is
sufficiently large, a new domain can be formed inside it by flipping spins
belonging to another sublattice, and so on.

A convenient method of describing the full set of ground states is based
on the notion of a zero-energy domain wall. \cite{K86} In the system under
consideration, a zero-energy domain wall can be defined as a line passing
through the bonds of the triangular lattice in such a way that each
segment of this line separates two plaquettes corresponding to different
regular states. This means that each segment of such a wall separates two
unparallel spins which on the original triangular lattice are
next-to-nearest neighbors and therefore belong to the same sublattice.
Then it is clear that the spin configuration in a ground state is uniquely
defined as soon as one specifies the orientations of three spins on an
arbitrary plaquette and the positions of all zero-energy domain walls. For
brevity, in the following we simply call these objects domain walls.

\begin{figure}[b]
\includegraphics[width=75mm]{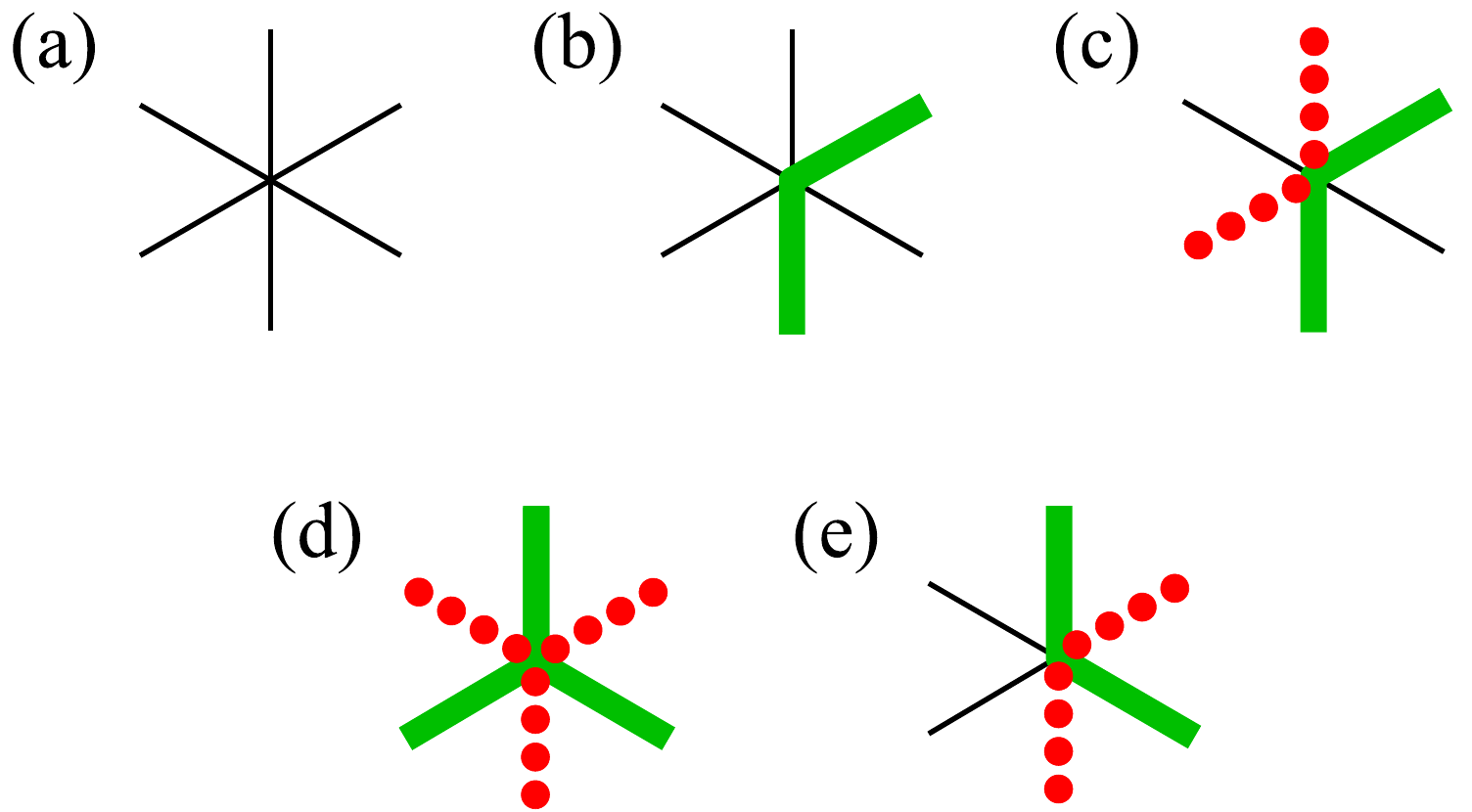}
\caption[fig-v] { (Color online) The loop representation: (a)-(c) allowed
configurations of loops, (d) additional vertex allowed only at
$\Phi=\arccos(-1/3)$, (e) the vertex corresponding to an intersection of
loops (allowed only at $\Phi=\pi/2$). The domain walls on different
sublattices are shown with lines of different types.} \label{fig-v}
\end{figure}

By definition, a regular state contains no domain walls, whereas flipping
of a single spin leads to the formation of the simplest closed domain wall
consisting of six segments joining each other at angle $2\pi/3$, i.e. a
hexagon. By flipping a larger number of spins belonging to the same
sublattice one can construct more complex domain walls. However, it is
clear that any wall has to be closed (or end at the boundary) and its
neighboring segments always have to form angles of $\pm 2\pi/3$ with
respect to each other, as in Fig. \ref{fig-v}(b).

Since each domain wall can be associated with spin-flipping on a
particular sublattice of the triangular lattice, all domain walls can be
partitioned into three classes (which below are called colors). The domain
walls of a given color live on one of the three equivalent honeycomb
sublattices into which the bonds of the triangular lattice can be
partitioned. Therefore, just by construction walls of the same color have
no possibility of crossing or touching each other.

\begin{figure}[t]
\includegraphics[width=85mm]{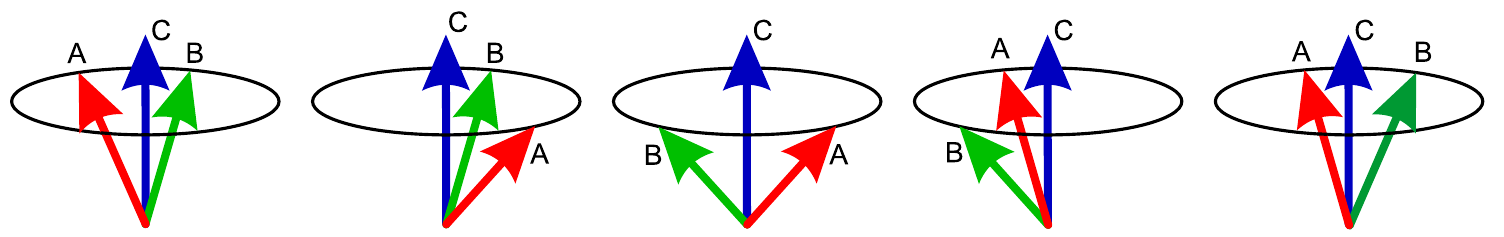}
\caption[fig-f] { (Color online) From left to right, configurations
obtained after successive flips of spins A, B, A, B. In the final
configuration, the triad of spins is rotated around C by an angle
$4\Psi(\Phi)$ with respect to the original one. } \label{fig-f}
\end{figure}

A less evident property is that for a generic value of $\Phi$ walls of
different colors are also unable to cross each other. This is so because
when going around an intersection of two walls [see Fig. \ref{fig-v}(e)]
one has to perform successive spin flips on sublattice $\alpha$ (where
$\alpha=A,B,C$), then on another sublattice $\beta$ and once again on
$\alpha$ and then on $\beta$, and return after that to the same
configuration. At a formal level, this corresponds to the fulfillment of
the condition
\begin{equation}                                             \label{RaRb}
    (\hat{R}_\beta\hat{R}_\alpha)^2 = 1\,,
\end{equation}
where $\hat{R}_\alpha$ 
denotes the operation of spin-flipping on sublattice $\alpha$
($\hat{R}^2_\alpha\equiv 1$). However, since each of the operators
$\hat{R}_\alpha$ and $\hat{R}_\beta$ involves a rotation of a spin around
the third spin of the triad by the angle $2\Psi(\Phi)$, where
\begin{equation}                                              \label{Psi}
\Psi(\Phi)=\arccos\left(\frac{\cos\Phi}{1+\cos\Phi}\right)\in\left(\pi/3,
\pi\right),
\end{equation}
the application of $(\hat{R}_\beta\hat{R}_\alpha)^2$ rotates the whole
configuration by $4\Psi(\Phi)$. It is easy to check that one returns to
the initial configuration only for $\Phi=\pi/2$ (when $\Psi$ is also equal
to $\pi/2$), whereas for all other values of $\Phi$ the operator
$(\hat{R}_\beta\hat{R}_\alpha)^2$ brings a triad of spins into a
configuration rotated with respect to the original one, as
illustrated in Fig. \ref{fig-f}. 

Condition (\ref{RaRb}) can be also rewritten as
$\hat{R}_\alpha\hat{R}_\beta = \hat{R}_\beta\hat{R}_\alpha$, which means
that the zero-energy domain walls can cross each other only when operators
$\hat{R}_\alpha$ commute with each other. However, walls of different
colors can touch each other, as shown in Fig. \ref{fig-v}(c).

As a consequence of the rules discussed above, for each site of the
triangular lattice there are only three possibilities: (i) the absence of
any domain wall passing through this site [see Fig. \ref{fig-v}(a)], (ii)
the presence of a single domain wall whose segments form an angle $2\pi/3$
with each other [see Fig. \ref{fig-v}(b)] and (iii) the presence of two
domain walls touching each other [see Fig. \ref{fig-v}(c)].
Therefore, to understand the zero-temperature properties of the system,
one has to study a loop model on the triangular lattice in which on each
site only the configurations shown in Figs. \ref{fig-v}(a)-(c) are allowed
(as well of course as those that can be obtained from them by rotation),
whereas all other configurations are prohibited [for example, the
intersection of two domain walls shown in Fig. \ref{fig-v}(e)]. The two
exceptions where other configurations of zero-energy domain walls are
allowed are the cases $\Phi=\arccos(-1/3)$ and $\Phi=\pi/2$, discussed
respectively in Sec. \ref{Potts} and Sec. \ref{Pi/2}.


One can introduce on each triangular plaquette $k$ a binary variable
$\chi_{k}=\pm 1$ defined in such a way that it is perfectly ordered in any
regular ground state and changes its sign on crossing each domain wall. In
terms of the original spin variables ${\bf S}_{j}$ the value of $\chi_{k}$
is given by the sign of the mixed product ${\bf S}_{{j}_A(k)} \cdot [{\bf
S}_{{j}_B({k})}\times{\bf S}_{{j}_C({k})}]$, where the three sites
${j}_A({k})$, ${j}_B({k})$ and ${j}_C({k})$ belonging to plaquette ${k}$
are labelled after the sublattice to which they belong. Below for
simplicity we call variable $\chi_{k}$ chirality of plaquette ${k}$,
although more accurately it should be called staggered chirality.

\section{Order-parameter manifold \label{O(n)}}

Let us define neighboring regular states as regular states which can be
transformed into each other by flipping the spins of one of the three
sublattices. From the definition of a domain wall it is clear that in any
ground state the domains lying on both sides of any domain wall belong to
neighboring regular states. Since each regular state has exactly three
neighboring states, at zero temperature the order-parameter manifold of
our system has the topology of an infinite Cayley tree (also known as
Bethe lattice) with coordination number $n=3$. Each node of this tree
represents a regular ground state, while the links connect neighboring
states. If one ascribes to each link on the tree the color of the
corresponding domain wall, the three links connecting any node with its
neighbors will all be of different colors.

\begin{figure}[b]
\includegraphics[width=70mm]{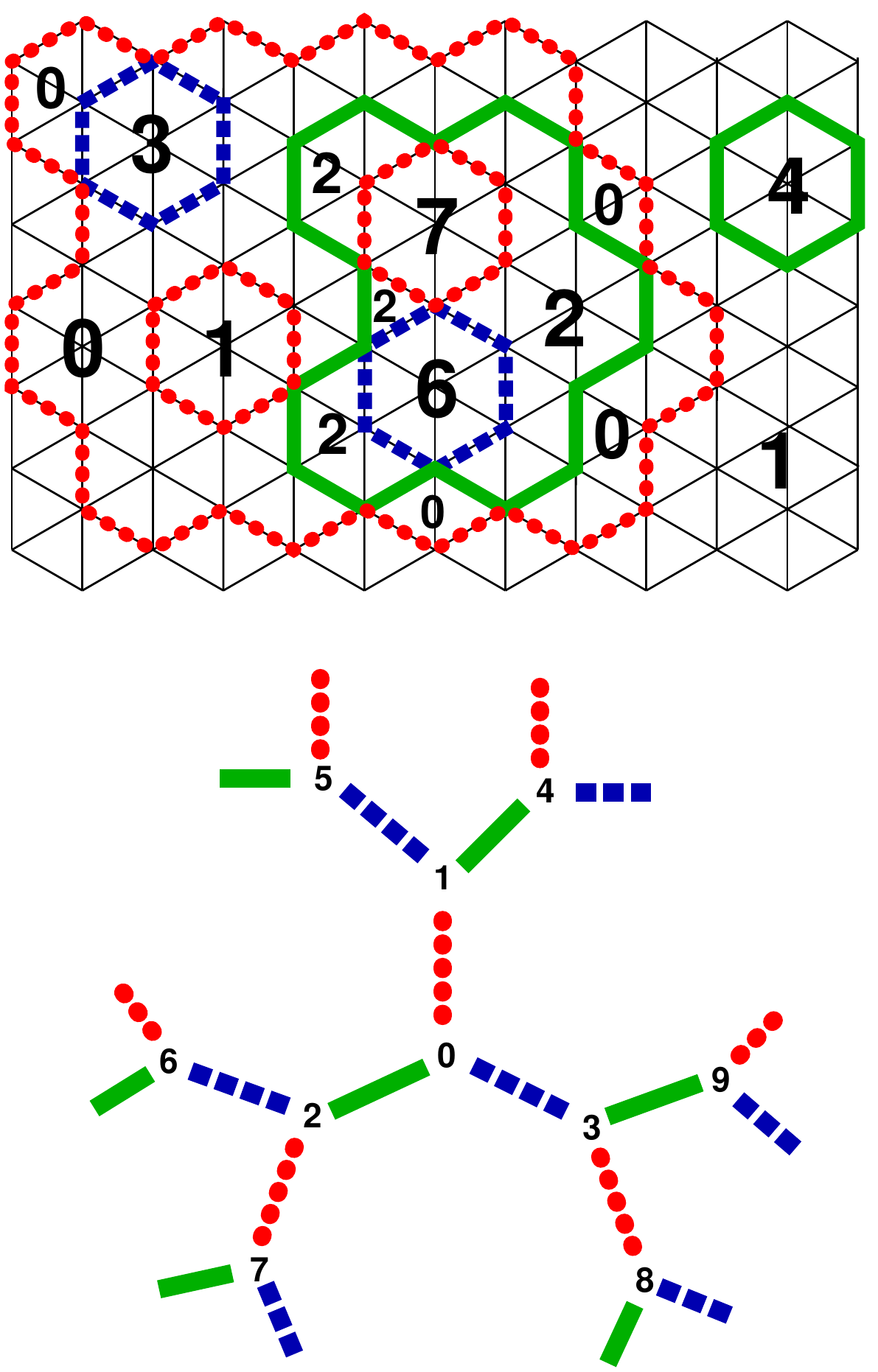}
\caption[fig-v] { (Color online) Top panel: Example of loop covering. The
numbers refer to the positions of different domains on the Cayley tree
shown in the lower panel. The numbering is arbitrary and has been included
to keep track of the relative positions on the tree.} \label{fig_loops}
\end{figure}

Fig. \ref{fig_loops} illustrates the correspondence between the domains of
various regular states and the nodes of the Cayley tree whose links are
colored in such a way. On the top panel, the domains corresponding to the
same regular state are marked by the same number (the numbering being
arbitrary). On the bottom panel, the same set of numbers is used to mark
the nodes of the Cayley tree. The distance on the tree corresponds to the
minimal number of spin flips one has to make (in other terms, the minimal
number of domain walls one has to cross) to get from one regular state to
another.

The tree-like topology of the order-parameter manifold (that is, the
absence of closed loops) is ensured by the impossibility (for a generic
value of $\Phi$) to return to the same state after crossing a sequence of
domain walls corresponding to a product of operators $\hat{R}_\alpha$
which cannot be reduced to unity by application of the identity ${\hat
R}_\alpha^2=1$. This property follows from the fact that intersections of
domain walls are not allowed.

Another model with the same structure of the order-parameter manifold is
obtained when one considers a loop model on the honeycomb lattice
\cite{DMNS,Nienh} in which each closed loop of length $L$ is ascribed a
factor
\begin{equation}                                             \label{w(L)}
w(L)=nK^{L}\,.
\end{equation}
In this model the loops cannot intersect or touch each other simply
because the geometry of the honeycomb lattice does not allow for this. The
number $n$ can then be interpreted as the number of different colors the
loops can have. This allows one to interpret these colored loops as domain
walls \cite{DMNS} separating the states on a Cayley tree with coordination
number $n$ by following the same rule that the states separated by a
domain wall of a given color correspond to neighboring nodes connected by
a link of the same color.

For $n=3$ the only difference between the two models is that in our
model the loops of different colors are living on three different
(interpenetrating) honeycomb lattices, whereas in the loop model of Ref.
\onlinecite{DMNS} 
they are living on the same honeycomb lattice. However in both models the
loops cannot cross each other or have an overlap, which ensures that the
two models have the same order-parameter manifold and therefore can be
expected to belong to the same universality class.

In Ref. \onlinecite{DMNS} it has been shown that the loop model defined by
Eq. (\ref{w(L)}) is a dual representation of a ferromagnetic $O(n)$ model
whose partition function can be written as
\begin{equation}                                              \label{O(N)}
Z_{O(n)}=\mbox{Tr}\left[\prod_{({kk'})}\left(1+K{\bf s}_k {\bf s}_{k'}
\right)\right]\,,
\end{equation}
where ${\bf s}_k$ are $n$-dimensional unit vectors defined at the sites
${k}$ of the honeycomb lattice, the product is taken over all bonds
$({kk'})$ of this lattice and the trace implies the unweighted integration
over all variables ${\bf s}_k$. On the other hand, it is also well known
that in $O(n)$ models with $n>2$ the long range order in ${\bf s}_k$ is
always destroyed by small continuous fluctuations of the order parameter
(the spin waves), \cite{Pol} which leads to an exponential decay of the
correlation functions $\langle{\bf s}_{k_1}{\bf s}_{k_2}\rangle$ with the
distance between ${k}_1$ and ${k}_2$.

It follows from the analysis of Refs. \onlinecite{EK} and \onlinecite{Sw}
that in terms of the loop representation
the exponential decay of $\langle {\bf s}_{k_1}{\bf s}_{k_2}\rangle$ means
that the insertion into the system of a string (an open loop) going from
${k}_1$ to ${k}_2$ leads to the suppression of the partition function of
the system by a factor which exponentially depends on the distance between
$k_1$ and $k_2$. This suggests that long loops have large effective free
energy and accordingly the probability of finding a sufficiently large
loop also decays exponentially with the size of this loop. As a
consequence, the typical number of irreducible loops separating the two
points cannot experience an unrestricted growth with the increase of the
distance between these points and has to saturate at some finite value.

This means that  in terms of the order parameter defined on a Cayley tree
the system is in the ordered phase, that is, it remains localized in some
region of the order parameter manifold even when its size tends to
infinity. From the universality we expect the loop model constructed in
the previous section (which has the same symmetry of the order-parameter
manifold) to be also in the ordered phase. In terms of original spin
variables ${\bf S}_j$, such a situation corresponds to the existence of a
genuine long-range order. This is especially evident for $\Phi\ll 1$. It
is clear that if in a typical ground state the system is localized on the
tree within the region of diameter $D_{\rm typ}$, then for $\Phi\ll
1/D_{\rm typ}$ the distant spins will be almost parallel to each other.

To support the hypothesis that there should be no qualitative difference
between the system in which the loops of different colors live on
different sublattices and the one where they live on the same sublattice,
we analyze in Appendix \ref{RProblem} a reduced problem with $n=2$. In
particular, we explicitly demonstrate that the system in which the
non-intersecting loops of two different colors live on two
interpenetrating honeycomb lattices can be in an exact way transformed
into the system in which the loops of two different colors live on the
same honeycomb lattice.

\section{Comparison with the four-state antiferromagnetic Potts model
         \label{Potts}}

When $\Phi=\arccos(-1/3)$, different spins ${\bf S}_{j}$ can have only
four different orientations in any ground state, which means that at zero
temperature the system is equivalent to the four-state AF Potts model. It
has been shown in Refs. \onlinecite{B70} and \onlinecite{MN} that the set
of ground states of the four-state AF Potts model on the triangular
lattice allows a mapping onto the set of ground states of the three-state
AF Potts model on a {\em kagom\'e} lattice, whose exact solution at zero
temperature was constructed by Baxter in 1970. \cite{B70} His results
imply that the four-state AF Potts model has a finite extensive entropy
per site whose numerical value is approximately equal to $0.379$.

Much later Huse and Rutenberg \cite{HR} have demonstrated that Baxter's
solution corresponds to an algebraic decay of spin correlations on the
same triangular sublattice of a {\em kagom\'e} lattice: $\langle{\bf
S}_{{j}_1}\cdot {\bf S}_{{j}_2}\rangle \propto 1/r_{12}^{\eta}$ with $\eta=4/3$,
$r_{12}$ being the distance between $j_1$ and $j_2$ (the application
of the same approach shows that in terms of the four-state AF Potts model
on the triangular lattice $\eta=1/3$). This conclusion was reached by
constructing another mapping \cite{HR,KH96} which puts each ground state
of the three-state AF Potts model on the {\em kagom\'e} lattice into
correspondence with a ground state of a solid-on-solid (SOS) model
describing fluctuations of a two-dimensional interface in a
four-dimensional space. In this SOS model height variables ${\bf u}_{j}$
(defined on the sites ${j}$ of the original triangular lattice ${\cal T}$)
are points on an auxiliary triangular lattice ${\cal T}_{*}$ defined in
the transverse (height) space. The only restriction on possible
configurations of ${\bf u}$ is that if ${j}$ and ${j'}$ are nearest
neighbors on $\cal T$, then ${\bf u}_j$ and ${\bf u}_{j'}$ have to be
nearest neighbors on ${\cal T}_{*}$.

To make the description of the system more transparent, it is convenient
to introduce the locally coarse-grained heights ${\bf
h}_k\equiv\frac{1}{3}[{\bf u}_{{j}_A({k})} +{\bf u}_{{j}_B({k})}+{\bf
u}_{{j}_C({k})}]$, that is the averages of variables ${\bf u}$ over the
three sites belonging to a given triangular plaquette of $\cal T$.
\cite{KH96,K02} These variables can be considered as defined at the sites
${k}$ of the honeycomb lattice $\cal H$ dual to $\cal T$, and their values
belong to the honeycomb lattice ${\cal H}_{*}$ dual to ${\cal T}_{*}$.
Variables ${\bf h}_k$ are more convenient than variables ${\bf u}_j$
because, in any regular state, all variables ${\bf h}_k$ are equal (in
contrast to variables ${\bf u}_j$). On the other hand, each domain wall
separates ${\bf h}_k$ and ${\bf h}_{k'}$ which are nearest neighbors on
${\cal H}_{*}$. Therefore, if ${k}$ and ${k'}$ are nearest neighbors on
$\cal H$, then ${\bf h}_k$ and ${\bf h}_{k'}$ have to be either equal or
nearest neighbors on ${\cal H}_{*}$.

Thus, for \makebox{$\Phi=\arccos(-1/3)$}, different regular states can be
associated with different points of ${\cal H}_{*}$, which means that the
order-parameter manifold instead of being a Cayley tree with coordination
number $n=3$ (as for generic values of $\Phi$) has the topology of a
periodic lattice with the same coordination number, namely, of the
honeycomb lattice. \cite{KH96,comment-24} On a formal level, this follows
from the existence of the identity $(\hat{R}_\beta\hat{R}_\alpha)^3\equiv
1$. This difference in the topology of the order-parameter manifold leads
to a change of the universality class, and, instead of being long-ranged,
the correlations are algebraic. In terms of the loop representation
introduced in Sec. \ref{GrStates}, a special feature of the case
$\Phi=\arccos(-1/3)$ is that in addition to vertices $a$, $b$ and $c$ (see
Fig. \ref{fig-v}), the vertex $d$ shown in the same figure is also allowed
(with the same unitary weight as all other vertices).

It is known from the analysis of Ref. \onlinecite{HR} that the
\makebox{(2+2)-dimensional} SOS model corresponding to the Potts model
under consideration is in the rough phase, where the large scale
fluctuations of ${\bf h}$ can be described by a free field effective
Hamiltonian,
\begin{equation}                                      \label{Heff}
     H_{\rm eff}=\frac{J}{2}\int_{}^{}d^2{\bf r}(\nabla{\bf h})^2\,,
\end{equation}
however, the value of the dimensionless rigidity modulus $J$ 
is such that the perturbation responsible for the discreteness of $\bf h$
is marginal. In other terms, the model is located exactly at the point of
the roughening transition. Therefore, any additional suppression of height
fluctuations should drive the system into the ordered phase. In
particular, we argue
below that this happens when one suppresses 
the formation of vertices $d$ (which for a generic value of $\Phi$ are
not allowed).

In terms of height variables ${\bf h}_{k}$ the main property of the loop
model defined by vertices $a$, $b$ and $c$ is that its order-parameter
manifold does not allow for the existence of closed loops in the height
space. As a consequence, when going around any of these vertices, the
vector ${\bf h}$ makes in the height space just a trivial loop that can be
contracted to a point and has zero area. In contrast to that, when going
around vertex $d$ the vector ${\bf h}$ sweeps a loop whose area is equal
to that of the elementary hexagon cell of ${\cal H}_{*}$. Accordingly, in
terms of height variables ${\bf h}_k$ an interaction that keeps the
weights of vertices $a$, $b$ and $c$ unchanged but leads to the partial or
complete suppression of vertices $d$ can be written as
\begin{equation}                                                 \label{Vd}
    V_d=\frac{J_d}{2}\sum_{{j}}^{}\Gamma^2_{j}\,,
\end{equation}
where $J_d>0$,
\begin{equation}                                                 \label{Sj}
    \Gamma_{j}=\frac{1}{2}\sum_{p=1}^{6}{\bf h}_{k_p(j)}
    \times{\bf h}_{k_{p+1}(j)}\,
\end{equation}
is the area of a loop swept by ${\bf h}$ when going counterclockwise
around the hexagonal plaquette of the dual lattice  surrounding
site ${j}$ of the original lattice, and $p$ numbers (in the same
direction) the six sites of the dual lattice belonging to this plaquette.
Naturally, complete suppression corresponds to $J_d\rightarrow+\infty$\,.

In the framework of a continuous description, Eq. (\ref{Vd}) should be
replaced by
\begin{equation}                                                 
    V_d=\frac{J_d}{2}\int_{}^{}d^2{\bf r}\left(
    \frac{\partial {\bf h}}{\partial r_1}\times
    \frac{\partial {\bf h}}{\partial r_2}\right)^2\,,
\end{equation}
where $r_1$ and $r_2$ are the two components of the vector ${\bf r}$ and
the quantity in brackets is the Jacobian of the mapping ${\bf
r}\rightarrow {\bf h}$. It seems natural to expect that the addition of
such an interaction with $J_d>0$ to the large-scale Hamiltonian
(\ref{Heff}) would lead to the suppression of the fluctuations of ${\bf
h}$ and indeed the perturbative treatment demonstrates that already in the
first order of the expansion in powers of $J_d$ one obtains a correction
to $J$.
For $J_d>0$ this correction is positive and therefore has to shift the
system into the ordered phase.  This provides one more argument in favor
of the conclusion that for a generic value of $\Phi$ the system has to be
in the ordered phase, which corresponds to the existence of a genuine
long-range order in terms of the original spin variables ${\bf S}_{j}$.

However, the existence of such a long-range order does not necessarily
imply the presence of the long range-order in chirality. Even if in a
typical configuration the system is localized in some region of the
order-parameter manifold, its distribution does not have to be centered on
a particular node of the order-parameter tree, but can also be, for
example, symmetric with respect to the center of a link connecting two
neighboring nodes. In such a case there will be no long-range order in
chirality, although the original spin variables ${\bf S}_{j}$ will be
ordered on all three sublattices.

\section{The case $\Phi=\pi/2$ \label{Pi/2}}

In the case $\Phi=\pi/2$, the spins 
on neighboring sites always have to be perpendicular to each other (where
``always'' means ``in any ground state''). As a consequence, each spin
always remains flippable independently of the flips made by other spins.
In such a situation it is natural to describe the ground states of the
system in terms of Ising pseudospins $\sigma_{j}=\pm 1$ such that ${\bf
S}_{j}={\bf S}^{\alpha({j})} \sigma_{j}$, where $\alpha({j})=A,B,C$
denotes the sublattice to which site ${j}$ belongs and ${\bf S}^A$, ${\bf
S}^B$, ${\bf S}^C$ are three unit vectors perpendicular to each other.
Since all possible sets of $\sigma_{j}$ are allowed, the residual entropy
per site is equal to $\ln 2\approx 0.693$ and the zero-temperature
spin-spin correlation function $\langle{\bf S}_i \cdot {\bf S}_j\rangle$
vanishes as soon as ${i}\neq {j}$.

However, the system has long-range order in terms of the spin-nematic
order parameter, $S^aS^b-\delta^{ab}/3$, \cite{andreev} with three
different easy axes (which are perpendicular to each other) for the three
sublattices. In Refs. \onlinecite{arikawa} and \onlinecite{LMP} devoted to
the investigation of model (\ref{eq:Hbb}) with $S=1$, the phase with such
a structure of the order parameter (realized for
$\pi/4\leq\theta\leq\pi/2$) was called the antiferroquadrupolar phase. The
difference between the quantum system with $S=1$ and the classical limit,
$S\to\infty$, is that for $S=1$ the order parameter on each sublattice is
like in an easy-plane spin nematic, whereas for $S\to\infty$ it is like in an
easy-axis spin nematic.

As it was already mentioned in Sec. \ref{GrStates}, in terms of the loop
representation the case $\Phi=\pi/2$ corresponds to the situation where
the intersections of loops [see Fig. \ref{fig-v}(e)] are not prohibited.
In such a case, the three subsystems of loops (on different sublattices)
are completely decoupled from each other.

\section{Conclusion\label{Concl}}

We have investigated the classical Heisenberg model on the triangular
lattice with a spin-canting interaction, {\em i. e.} an interaction such
that the energy of a pair of spins is  minimized when the angle between
them is equal to \makebox{$\Phi \in (0,\pi)$}. In particular, the
interaction of two spins has this property if the antiferromagnetic
biquadratic exchange is of comparable strength with the bilinear exchange.
\cite{KY} Our analysis is focused on the case $\Phi \in (0,2\pi/3)$, where
the energy can be minimized simultaneously for all bonds. The family of
ground states is then characterized by a well-developed degeneracy
corresponding to an extensive residual entropy, which makes the question
whether at zero temperature the system is ordered, disordered or critical
a non-trivial one.

After constructing a complete classification of the ground states of the
model, we have demonstrated that at zero temperature its order-parameter
manifold has the structure of an infinite Cayley tree with coordination
number $n=3$ and therefore is isomorphic to that of the $O(n)$ model with
$n=3$. This suggests that the two models belong to the same universality
class and therefore the system under consideration should be, like the
$O(n=3)$ model, in the ordered phase, which means long-range order in
terms of the original spin variables ${\bf S}_{j}$. This conclusion has
been confirmed by showing that our model can be constructed by imposing an
additional restriction onto the four-state AF Potts model on the
triangular lattice, which leads to the suppression of the fluctuations of
the free field describing the large-scale fluctuations of this Potts model
and therefore pushes it from the point of roughening transition into the
ordered phase.

Numerical simulations of the classical bilinear-biquadrartic Heisenberg
model on the triangular lattice demonstrate \cite{KY} that at $J_2/J_1=-3$
and very low temperatures the spins in a typical configuration form a
non-collinear structure analogous to what can be expected from a typical
ground state in the presence of weak continuous fluctuations. However, the
authors of Ref. \onlinecite{KY} have not studied how the spin-spin
correlation functions behave in the $T\to 0$ limit (at any finite
temperature the long-range order in spin orientation in any case has to be
destroyed by the spin waves \cite{Pol,friedan}), so their results cannot
be used for checking our conclusions. Moreover, it still remains to be
elucidated if the behavior of the system at very low temperatures is
determined mostly by its zero-temperature properties, or whether the
removal of the accidental degeneracy by continuous fluctuations (spin
waves) can play an important role, as in the case of a Heisenberg or XY
antiferromagnet on the triangular lattice in the presence of an external
magnetic field. \cite{Kaw} Note that the only possibility for the
non-collinear phase considered in this paper to have a genuine long-range
order at finite temperatures consists in having a long-range order in
chirality, and our conclusions allow for the realization of such a
scenario.

The two exceptions from the generic behavior described above are the cases
$\Phi=\pi/2$ and \makebox{$\Phi=\arccos(-1/3)$} for which the model at
zero temperature is exactly solvable. For $\Phi=\pi/2$ the fluctuations of
any pair of spins are uncorrelated, $\langle{\bf S}_{i}\cdot{\bf
S}_{j}\rangle=0$, but the system has long range order in terms of the
spin-nematic order parameter 
with three different easy axes for the three sublattices (the
antiferroquadrupolar phase\cite{arikawa,LMP}). On the other hand, for
$\Phi=\arccos(-1/3)$ the model is equivalent to the four-state AF Potts
model, in which the chirality and spin correlations are algebraic,
$\langle{\chi}_{{k}_1}{\chi_k}_{{k}_2}\rangle \propto 1/r_{12}^{4}$
\makebox{(see Ref. \onlinecite{HR}),} $\langle{\bf S}_{{j}_1}\cdot {\bf
S}_{{j}_2}\rangle \propto 1/r_{12}^{1/3}$.

\begin{acknowledgments}
We acknowledge useful discussions with Sandro Wenzel.  This work was
supported by the Swiss National Fund, by MaNEP and by the Hungarian OTKA
Grant No. K73455.
S.E.K. and K.P. are grateful for the hospitality of the EPFL
where most of the work has been completed.
\end{acknowledgments}

\appendix

\section{Reduced problem\label{RProblem}}

Let us consider a ``reduced'' loop problem which differs from the one
formulated in Sec. \ref{GrStates} by the suppression of the formation of
loops on one of the three honeycomb sublattices. This is equivalent to
saying that the spins on one of the three triangular sublattices are not
allowed to flip and remain aligned in the same direction (below this
sublattice is denoted ${\cal T}_C$ and two other sublattices ${\cal T}_A$
and ${\cal T}_B$). In such a situation, the formation of loops on the
honeycomb lattice ${\cal H}_{AB}$ 
formed by the sites of ${\cal T}_A$ and ${\cal T}_B$ is impossible and the
loop system consists of only two types of loops (on honeycomb lattices
${\cal H}_{AC}$ and ${\cal H}_{BC}$).

When all spins on ${\cal T}_C$ point  in the same direction, ${\bf
S}^C_{j}={\bf S}^C$, the orientation of spins on the two other sublattices
(${\cal T}_A$ and ${\cal T}_B$) can be described by a single variable
$\psi$, the angle between the projection of ${\bf S}_{j}$ on the plane
perpendicular to ${\bf S}^{C}$ and some reference direction in this plane.
Since the angle between neighboring spins is always equal to $\Phi$, the
difference between the values of $\psi$ on neighboring sites of ${\cal
H}_{AB}$ has to be equal (modulo $2\pi$) to $\pm\Psi(\Phi)$, where
$\Psi(\Phi)$ 
is given by Eq. (\ref{Psi}). For $\Phi\in(0,2\pi/3)$, $\Psi(\Phi)$ belongs
to the interval $(\pi/3,\pi)$ and is a monotonically increasing function
of $\Phi$.

For $\Psi(\Phi)\neq\pi/2$ (that is, $\Phi\neq\pi/2$), the existence of
such a relation between the orientations of neighboring spins allows one
to introduce integer variables $u_{j}$ defined on the sites of honeycomb
lattice ${\cal H}_{AB}$
in such a way that on neighboring sites of ${\cal H}_{AB}$ (belonging to
different sublattices) they always differ by $\pm 1$. In particular, one
can choose variables $u_{j}$ to be even on ${\cal T}_A$ and odd on ${\cal
T}_B$. Then on neighboring sites of the same triangular sublattice
these variables either differ by $\pm 2$ (when these sites are separated
by a domain wall) or are equal to each other (in the absence of such a
wall).

Accordingly, the system of $AC$ and $BC$ loops can be discussed in terms
of a $(2+1)$-dimensional solid-on-solid (SOS) model defined on ${\cal
H}_{AB}$. In the SOS representation, variables $u_{j}$
have the meaning of surface heights, 
whereas each domain wall plays the role of a step between the regions in
which the values of either ${u}^A$ or of ${u}^B$ differ by $\pm 2$. The
SOS model defined in such a way can be considered as a generalization of
the BCSOS model (introduced by van Beijeren \cite{vB} on the square
lattice) to the honeycomb lattice.  In particular, the representation of
the model in terms of Ising-type pseudospins \makebox{$\sigma_{j}=\pm 1$}
can be constructed with the help of the same relations as for the BCSOS
model, \cite{Knops}
\begin{subequations}                                     \label{sigma}
\begin{eqnarray}
\sigma^A_{j} & = & \cos [(\pi/2){u}^A_{j}]\,,  \label{sigmaa}\\
\sigma^B_{j} & = & \sin [(\pi/2){u}^B_{j}]\,,  \label{sigmab}
\end{eqnarray}
\end{subequations}
where we still follow the convention that variables ${u}_{j}$ are even on
${\cal T}_A$ and odd on ${\cal T}_B$. The pseudospin variables
$\sigma_{j}$ are defined in such a way that each flip of the original spin
$S_{j}$ leads to a change of sign of $\sigma_{j}$. Accordingly,
pseudospins of opposite signs belonging to the same sublattice are
separated by domain walls.

All configurations allowed in this SOS model enter the partition function
of the reduced problem with the same weight (like in the infinite
temperature limit of the \makebox{BCSOS} model). The existence of the
mapping to a SOS model 
confirms that the suppression of spin flips on one of the three
sublattices reduces the order parameter manifold of the system from a tree
with coordination number $n=3$ to a trivial tree with coordination number
$n=2$.

The SOS model on ${\cal H}_{AB}$ introduced above can be further mapped
onto the 20-vertex model on the triangular lattice introduced by Baxter.
\cite{B69} In order to perform the mapping, one should put on each bond of
${\cal T}_C$ an arrow directed in such a way that the larger of the two
heights at the ends of the $AB$ bond crossed by this arrow is always to
the right of the arrow. Then on each site of ${\cal T}_C$ one will have
three incoming and three outcoming arrows (the ice rule) which precisely
corresponds to the selection of the allowed vertices in the 20-vertex model.

The exact solution of the 20-vertex model on the triangular lattice 
found by Baxter \cite{B69} demonstrates that when all vertices have the
same weight (as in our case) the dimensionless free energy per
vertex is equal to $-\frac{1}{2}\ln\frac{27}{4}$. In terms of the current
discussion this means that the residual entropy of the reduced problem
(per site of the original triangular lattice ${\cal T}$) is equal to
$\frac{1}{6}\ln\frac{27}{4}\approx 0.318$. The residual entropy
of the full loop problem should be larger than this value but lower than
that of the four-state AF Potts model (approximately equal to $0.379~$
\cite{B70}).

Let us now compare the contributions to the partition function of the
reduced problem which can be associated with different configurations of
the $BC$ loops (in other terms, with different allowed configurations of
variables $u_{j}^A$). The contribution from the configuration in which the
$BC$ steps are completely absent (that is, all variables $u_{j}^A$ are
equal to each other, $u_{j}^A=u^A$) can be found very easily. In such a
situation, all variables $u_{j}^B$ can acquire two values $u^B=u^A\pm 1$
independently of each other. Therefore, the corresponding contribution is
equal to $2^{N_0}$, where $N_0$ is the number of sites on each of the
three sublattices.

If there is a closed $BC$ loop of length $L$
separating the regions in which the values of the variables $u_{j}^A$
differ by 2, then on all $L/2$ sites from ${\cal T}_B$ passed by this loop
the variables $u_{j}^B$ cannot fluctuate and have to be equal to the
average between the values of $u^A$ on both sides of the loop. This means
that the presence of a loop of length $L$ decreases the number of allowed
configurations on sublattice ${\cal T}_B$ from $2^{N_0}$ to
$2^{{N_0}-L/2+1}$. The additional factor of 2 appears because for each
loop configuration this loop can represent two types of steps (positive
and negative).

After noting that an analogous reduction is induced by every loop on
${\cal H}_{BC}$, one can conclude that after summation over all allowed
configurations of variables $u_{j}^B$ one obtains an SOS model defined
on the triangular lattice ${\cal T}_A$. In this model 
the values of height variables on neighboring sites of ${\cal T}_A$ can
either be equal or differ by $\pm 2$, whereas each step is ascribed a
weightfactor exponentially decaying with its length. This model can also
be interpreted \cite{DMNS, SD} as the Ashkin-Teller model on the
triangular lattice \cite{Wu77} with one of the weights equal to zero.

Thus we have reduced a model in which the loops of two different types are
living on different interpenetrating honeycomb lattices to a model in
which all loops are living on the same honeycomb lattice. However the
weight factor corresponding to each loop
\begin{equation}                                             \label{w(2L)}
w(L)=2(1/\sqrt{2})^{L}\,.
\end{equation}
contains a factor $2$ related to the existence of two types of loops.
The loop model defined by the weight factor (\ref{w(2L)}) is just a
particular case of the more general model \cite{DMNS} with $w(L)$ given by
Eq. (\ref{w(L)}). For $-2\leq n\leq 2$ this model is known \cite{Nienh} to
have a phase transition at \makebox{$K_c(n)=1/\sqrt{2+\sqrt{2-n}}$}.

This suggests that for $n=2$ and \makebox{$K=1/\sqrt{2}$} the SOS model on
${\cal T}_A$ defined two paragraphs above is located exactly at the point
of the roughening transition, where large-scale fluctuations of ${u}^A$
diverge logarithmically and can be described by a (dimensionless)
free-field effective Hamiltonian,
\begin{equation}                                      \label{Heff-b}
     H_{\rm eff}=\frac{J}{2}\int_{}^{}d^2{\bf r}(\nabla{{u^A}})^2\,,
\end{equation}
with dimensionless rigidity modulus $J=\pi/8$. This value of $J$ ensures
the marginality of the operator $-\cos(\pi u^A)$ favoring even values of
$u^A$. It follows then from Eq. (\ref{sigmaa}) that the correlation
function \makebox{$\langle\sigma^{A}_{{j}_1}\sigma^{A}_{{j}_2}\rangle$}
decays algebraically,
\makebox{$\langle\sigma^{A}_{{j}_1}\sigma^{A}_{{j}_2}\rangle \propto
1/r_{12}^{\eta_\sigma}$} with $\eta_\sigma=1$, $r_{12}$ being the distance
between $j_1$ and $j_2$.

It is clear already from symmetry that the analogous correlation function
on ${\cal T}_B$ has to decay in exactly the same way as on ${\cal T}_A$.
The law of this decay can also be derived with the help of another
approach. It can be rigorously shown that the correlation function of the
two pseudospins on sublattice ${\cal T}_B$ is equal to the probability
that they belong to the same $BC$ loop. In terms of SOS representation
such a loop corresponds to  a closed step at which variables $u^A_{j}$
jumps by $\pm 2$.

In the framework of the SOS model defined on sublattice ${\cal T}_A$,  the
total statistical weight of configurations in which the two given sites on
${\cal T}_B$ belong to the same $BC$ loop can be found by reversing the
sign of the step on one of the two segments into which these two sites
split the loop. \cite{SaD,KH96} This transforms the full set of such
configurations into the set of configurations allowed in a system with a
neutral pair of screw dislocations on going around each of which $u^A_{j}$
changes by $b=\pm 4$. When the large-scale fluctuations of $u^A$ can be
described by the Hamiltonian (\ref{Heff-b}), the dimensionless free energy
associated with such a dislocation pair will be given by $F\approx
(Jb^2/2\pi)\ln r_{12}$, which shows that
\makebox{$\langle\sigma^B_{{j}_1}\sigma^B_{{j}_2} \rangle\propto
r_{12}^{-1}$}. The same value of the exponent follows also from the
possibility to unambiguously interpret the steps in the considered model
as the contours of equal height. \cite{KH95} The consistency of the
results produced by different approaches supplies an additional
confirmation that the value of the rigidity modulus $J$ in Eq.
(\ref{Heff-b}) has been correctly chosen.

In the framework of the reduced problem, the expression for the chirality
of a plaquette is reduced to the product of two pseudospins on neighboring
sites, $\chi_{k}=\sigma_{{j}_A({k})}\sigma_{{j}_B({k})}$. In order to find
how chirality correlations decay with distance, it is convenient to
express this product in terms of locally coarse-grained height
$h_{k}=\frac{1}{2}[u_{{j}_A({k})}+u_{{j}_B({k})}]$ as
\begin{equation}                                             \label{chi}
 \chi_{k}=\sigma_{{j}_A({k})}\sigma_{{j}_B({k})}=\sin(\pi h_{k})\,.
\end{equation}
The variables $h_{k}$ are defined on $AB$ bonds and acquire values which
are shifted from integers by $1/2$. It follows from the definition of
$h_{k}$ that the values of these variables on adjacent bonds can be either
equal to each other or differ by $\pm 1$.
In particular, in any regular state the values of 
$h_{k}$ are the same on all $AB$ bonds. Since the large-scale fluctuations
of $h$ have to be described by the same Hamiltonian, Eq. (\ref{Heff-b}),
as those of $u^A$, it follows from Eq. (\ref{chi}) that chirality
correlations have to decay with exponent $\eta_\chi=4$ (as in the
four-state antiferomagnetic Potts model, see Sec. \ref{Potts}). This means
that $\chi$ is a marginal operator.

Note that the decay of $\langle\chi_{{k}}\chi_{{k}'}\rangle$ is much
faster than that of the product of
$\langle\sigma_{{j}_A({k})}\sigma_{{j}_A({k}')}\rangle$ and
$\langle\sigma_{{j}_B({k})}\sigma_{{j}_B({k}')}\rangle$. In terms of the
loop representation the origin of this property is quite clear. The
presence of an $AC$ loop passing through the sites ${j}_A({k})$ and
${j}_A({k}')$ strongly decreases the number of configurations with a $BC$
loop passing through the sites ${j}_B({k})$ and ${j}_B({k}')$. From the
analogy with the case of a single loop (discussed two paragraphs above),
the fraction of configurations in which two such loops are present
simultaneously can be estimated by considering a system with a neutral
pair of screw dislocations with $b=\pm 8$, which again gives
$\eta_\chi=4$.


\begin{thebibliography}{99}
\bibitem{nakatsuji} S. Nakatsuji et al., Science {\bf 309}, 1697 (2005).
\bibitem{haldane} F. D. M. Haldane, Phys. Lett. A {\bf 93}, 464 (1983);
Phys. Rev. Lett. {\bf 50}, 1153 (1983).
\bibitem{PRL_AKLT} I. Affleck, T. Kennedy, E. H. Lieb, and H. Tasaki,
Phys. Rev. Lett. {\bf 59}, 799 (1987).
\bibitem{fath1991} G. F\'ath and J. S\'olyom, Phys. Rev. B {\bf 44}, 11836
(1991).
\bibitem{Schollwock1996} U. Schollw\"ock, T. Jolicoeur, and T. Garel,
Phys. Rev. B {\bf 53}, 3304 (1996).
\bibitem{parkinson} J. Parkinson, J. Phys.: Condens. Matter {\bf 1}, 6709 (1989).
\bibitem{okunishi1} K. Okunishi, Y. Hieida, and Y. Akutsu,
Phys. Rev. B {\bf 60}, 6953(R) (1999); B {\bf 59}, 6806 (1999).
\bibitem{fath} G. F\'ath and P. B. Littlewood, Phys. Rev. B {\bf 58}, 14709(R) (1998).
\bibitem{manmana} S. R. Manmana, A. M. L\"{a}uchli, F. H. L. Essler, and F. Mila,
Phys. Rev. B {\bf 83}, 184433 (2011).
\bibitem{kawashima} K. Harada and N. Kawashima, Phys. Rev. B {\bf 65}, 052403 (2002).
\bibitem{arikawa} H. Tsunetsugu and M. Arikawa, J. Phys. Soc. Jap. {\bf 75}, 083701 (2006).
\bibitem{LMP}    A. L\"{a}uchli, F. Mila, and K. Penc,
                 Phys. Rev. Lett. {\bf 97}, 087205 (2006).
\bibitem{toth}   T. Toth, A. L\"{a}uchli, F. Mila, and K. Penc,
                 unpublished (arXiv:1110.2495).
\bibitem{LGT}    D.~H. Lee, G. Grinstein, and J. Toner,
                 Phys. Rev. Lett. {\bf 56}, 2318 (1986).
\bibitem{KY}     H. Kawamura and A. Yamamoto,
                 J. Phys. Soc. Jpn, {\bf 76}, 073704 (2007).
\bibitem{Suto}   A. S\"{u}t\"{o}, Z. Phys. B {\bf 44}, 121 (1981).
\bibitem{KB}     L.~P. Kadanoff and A. Brown,
                 Ann. Phys. (N.Y.) {\bf 121}, 318 (1979).
\bibitem{St}     G.~H. Wannier, Phys. Rev. {\bf 79}, 357 (1950);
                 Phys. Rev. B {\bf 7}, 5017E (1973);
                 R.~M.~F. Houtappel, Physica {\bf 16}, 425 (1950);
                 J. Stephenson, J. Math. Phys. {\bf 5}, 1009 (1964);
                                             {\bf 11}, 413 (1970).
\bibitem{B70}    R.~J. Baxter, J. Math. Phys. {\bf 11}, 784 (1970).
\bibitem{HR}     D.~A. Huse and A. D. Rutenberg,
                 Phys. Rev. B {\bf 45}, 7536 (1992).
\bibitem{KSS}    R. Koteck\'{y}, J. Salas, and A.~D. Sokal,
                 Phys. Rev. Lett. {\bf 101}, 030601 (2008).
\bibitem{K86}    S.~E. Korshunov, J. Stat. Phys. {\bf 43}, 17 (1986);
                 S.~E. Korshunov, A. Vallat, and H. Beck,
                 Phys. Rev. B {\bf 51}, 3071 (1995).
\bibitem{DMNS}   E. Domany, D. Mukamel, B. Nienhuis, and A. Schwimmer,
                 Nucl. Phys. {\bf B190} [{\bf FS3}], 279 (1981).
\bibitem{Nienh}  B. Nienhuis, Phys. Rev. Lett. {\bf 49}, 1062 (1982).
\bibitem{Pol}    A.~M. Polyakov, Phys. Lett. {\bf 59B}, 79 (1975).
\bibitem{EK}     J.~P. van der Eerden and H.~J.~F. Knops,
                 Phys. Lett. {\bf 66A}, 334 (1978).
\bibitem{Sw}
                 R.~H. Swendsen, Phys. Rev. B {\bf 17}, 3710 (1978).
\bibitem{MN}     C. Moore and M. E. J. Newman,
                 J. Stat. Phys. {\bf 99}, 630 (2000).
\bibitem{KH96}   J. Kondev and C.~L. Henley,
                 Nucl. Phys.  {\bf B464}, 540 (1996).
\bibitem{comment-24} Stricltly speaking, the ground states of the AF Potts model
                 have only 24-fold degenercy which corresponds to the
                 compatification of the above-mentioned honeycomb lattice
                 on a torus. However, in all ground states the winding numbers
                 obtained when going around any plaquette
                 cannot be nonzero, as a consequence of which
                 the system behaves itself at zero temperature as if its
                 order parameter space was an infinite honeycomb lattice.
\bibitem{K02}    S.~E. Korshunov,
                 Phys. Rev. B {\bf 65}, 054416 (2002);
                 Usp. Fiz. Nauk {\bf 176}, 233 (2006)
                 [Physics - Uspekhi {\bf 49}, 225 (2006)].
\bibitem{andreev}  A. F. Andreev and I. A. Grishchuk, Zh. Eksp. Teor. Fiz.
                 {\bf 87}, 467 (1984) [Sov. Phys. - JETP 60, 267 (1984)];
                 A.~V. Chubukov, J. Phys. Condens. Matter {\bf 2}, 1593 (1990).
\bibitem{friedan} D.~H. Friedan, Ann. Phys. (N.Y.) {\bf 163}, 31 (1983);
                 P. Azaria, B. Delamotte, T. Jolicoeur, and D. Mouhanna,
                 Phys. Rev. B {\bf 45}, 12612 (1992).
\bibitem{Kaw}    H. Kawamura, J. Phys. Soc. Jpn. {\bf 53}, 2452 (1984);
                 H. Kawamura and S. Miyashita, J. Phys. Soc. Jpn. {\bf 54},
                 4530 (1985); S.~E. Korshunov, Pis'ma ZhETF, {\bf 41}, 525 (1985)
                 [JETP Lett. {\bf 41}, 641 (1985)];
                 J. Phys. C {\bf 19}, 5927 (1986).
\bibitem{vB}     H. van Beijeren, Phys. Rev. Lett. {\bf 38}, 993 (1977).
\bibitem{Knops}  H.~J.~F. Knops, Ann. Phys. (N.Y.) {\bf 128}, 448 (1980).
\bibitem{B69}    R.~J. Baxter, J. Math. Phys. {\bf 10}, 1211 (1969).
\bibitem{SD}     Y. Stavans and E. Domany, Phys. Rev. B {\bf 27}, 3043 (1983).
\bibitem{Wu77}   F.~Y. Wu, J. Phys. C {\bf 10}, L23 (1977).
\bibitem{SaD}    H. Saleur and B. Duplantier,
                 Phys. Rev. Lett. {\bf 58}, 2325 (1987).
\bibitem{KH95}   J. Kondev and C.~L. Henley,
                 Phys. Rev. Lett. {\bf 74}, 4580 (1995).
\end{thebibliography}
\end{document}